\documentclass[aps,prc,twocolumn,nofootinbib,superscriptaddress]{revtex4-1}

\usepackage{graphicx}
\usepackage{amsmath}
\usepackage{amsfonts}
\usepackage{amsbsy}
\usepackage{bm}
\usepackage{xcolor}
\usepackage{hyperref}
\usepackage{float}
\usepackage{xspace}

\hypersetup{
    colorlinks,
    linkcolor={red!50!black},
    citecolor={blue!50!black},
    urlcolor={blue!80!black}
}

\newcommand{\beq}{\begin{equation}}
\newcommand{\eeq}{\end{equation}}
\newcommand{\bea}{\begin{eqnarray}}
\newcommand{\eea}{\end{eqnarray}}

\newcommand{\rvec}{{\bf r}}
\newcommand{\kvec}{{\bf k}}
\newcommand{\qvec}{{\bf q}}

\begin{document}

\title{Two-Body Scattering Observables from Finite-Volume Real-Time Evolution}

\author{Lucas Platter}
\affiliation{Department of Physics and Astronomy, University of Tennessee, Knoxville, Tennessee 37996, USA}
\affiliation{Physics Division, Oak Ridge National Laboratory, Oak Ridge, Tennessee 37831, USA}

\date{\today}

\begin{abstract}
  We study two-body scattering observables from real-time evolution in
  a finite periodic box. The system consists of two distinguishable
  particles on a two-dimensional lattice interacting through pointlike
  $s$- and $p$-wave interactions. We evolve their wave packets in real
  time, define detector observables through angular wedges in the
  relative coordinate, and attach infinite-volume labels obtained from
  the bound-state pole equation and the low-energy scattering
  amplitude. We train a convolutional neural network on this data and
  test its performance on held-out scattering problems and find that
  is able to predict the total magnitude and angular shape for
  previously unseen Hamiltonians.
\end{abstract}

\maketitle
\section{Introduction}
\label{sec:intro}
Quantum computers are especially well-suited to simulate real-time
evolution, since quantum circuits are composed of unitary gates and
thus naturally realize unitary
dynamics~\cite{Feynman:1981tf,Jordan:2012wd}. In nuclear theory,
time-evolution methods are well established in mean-field reaction
dynamics, for example in time-dependent Hartree--Fock studies of
collective motion and heavy-ion collisions~\cite{Simenel:2012ca}. For
lighter systems, various time-independent approaches describe
scattering and reaction processes
exactly~\cite{Kamada:2001tv}. However, the complexity of such
approaches grows rapidly with the number of
nucleons~\cite{Deltuva:2009zz,Gloeckle:1996zz}. Real-time evolution
offers a complementary route that avoids complications such as the
analytic structure of the scattering kernel and may extend more
naturally to larger systems.

Finite lattices representing a finite volume are a standard choice on
quantum devices. An important question is therefore how much
infinite-volume scattering information can be recovered from
finite-volume real-time observables. This program is complementary to
the L\"uscher finite-volume
approach~\cite{Luscher:1985dn,Luscher:1986pf}, in which
infinite-volume scattering information is inferred from the volume
dependence of the discrete energy spectrum. Here we use real-time
dynamical observables instead, which gives direct access to angular
information without requiring an explicit spectral
decomposition. Recent work has shown that, for gapped theories, the
corresponding finite-volume corrections are exponentially
suppressed~\cite{Burbano:2025reso}. Scattering observables have also
previously been extracted on quantum devices and in closely related
hybrid workflows~\cite{Sharma:2024bjs,Turro:2024vpc,Yusf:2025eqc,
  Bennewitz:2025mss}. 

In this work, we carry out the calculation in the simplest nontrivial
setting: two distinguishable particles (e.g. two neutrons in different
spin states) on a two-dimensional periodic lattice. The
two-dimensional setting is chosen as a proof of principle: it keeps
the Hilbert space manageable and makes the angular structure easy to
visualize. We evolve initialized wave packets, define detector
observables through angular wedges in the relative coordinate, study
the box-size dependence of the resulting signals, and attach
infinite-volume scattering labels computed within the same lattice
model to the collected data. Here detector observables are not
measurements by a physical apparatus, but lattice sums of the
relative-coordinate probability density over angular sectors.

In the standard continuum formulation, scattering observables are
defined from asymptotic plane-wave states and the associated
large-distance or large-time behavior of the wave
function~\cite{Taylor:1972pty}. The present wave-packet construction
is designed to approximate that situation inside a finite box: once
the packets have propagated far enough from the interaction region,
their angular redistribution retains the relevant asymptotic
information. In the future, such an approach could be combined with
quantum-computing methods for few-body scattering with composite
projectiles and targets~\cite{Chai:2025fwp,Davoudi:2024swh}.

We use a convolutional neural network (CNN) to map the finite-volume
wedge-time signal onto the corresponding infinite-volume scattering
amplitude. The network is trained on infinite volume total and
differential cross sections that are computed from the same lattice
Hamiltonian, ensuring internal consistency. Because the wedge
observables require only standard-basis measurements, {\it i.e.}
position-space probability sums rather than off-diagonal overlaps, the
approach is well suited to near-term quantum hardware, where
ancilla-free measurement circuits are strongly preferred.

Wave-packet scattering, finite-volume scattering analysis, and
quantum-computing approaches to scattering all have important prior
developments~\cite{Jordan:2012wd,Sharma:2024bjs,Turro:2024vpc,Yusf:2025eqc,Chai:2025fwp,Davoudi:2024swh,Luscher:1985dn,Luscher:1986pf}. Rule
and Stetcu have recently proposed extracting $S$-matrix elements from
time-dependent wave-packet overlaps on a quantum
device~\cite{Rule:2026brk}. That approach gives direct access to the
scattering amplitude at fixed energy but requires controlled
time-evolution circuits and ancilla qubits for each overlap
measurement. The present method trades that rigor for simultaneous
angular coverage and simpler measurements: all wedge bins are obtained
from a single time evolution followed by standard-basis readout.

The novelty of the present approach lies in the combination of these
ingredients into a single framework: finite-volume real-time lattice
evolution, detector observables defined by direct sums over lattice
points in relative-coordinate wedges, matched infinite-volume labels,
and supervised inference of scattering information from the
finite-volume dynamical signal.

The manuscript is organized as follows.  Section~\ref{sec:framework}
defines the lattice Hamiltonian and the interactions used in the
benchmarks, and derives the infinite-volume scattering amplitudes that
serve as supervised labels.  Section~\ref{sec:time-evolution}
describes the wave-packet initialization and real-time evolution.  We
introduce the wedge observables and their connection to the cross
section in Sec.~\ref{sec:observables}.  Section~\ref{sec:results}
presents the numerical benchmarks and machine-learning results.  We
conclude in Section~\ref{sec:summary} with a summary and an outlook.

\section{Two-body physics on a two-dimensional lattice}
\label{sec:framework}

We consider two distinguishable particles of equal mass
$m = 938.918$~MeV on a square lattice with spacing $a$ and periodic
boundary conditions. The lattice has $N\times N$ sites, so that the
box size is $L=Na$.  In the present numerical implementation we use a
small number of lattice spacings and a combination of attractive $s$-
and $p$-wave interactions.  Throughout this work we use
$\hbar=c=1$. We write the full Hamiltonian for our model as
\begin{align}
    H= T + V_{s} + V_p~, 
\end{align}
where the free Hamiltonian is
\begin{multline}
    T= \sum_{\bf n, \sigma}
    \frac{1}{2m a^2}\bigl(
    4c^\dagger_{{\bf n},\sigma}c_{{\bf n},\sigma}
    -\sum_{\mu=\pm \hat{x},\pm\hat{y}}c^\dagger_{{\bf n}+\mu,\sigma}c_{{\bf n},\sigma}
    \bigr)~,
\end{multline}
and $c_{\mathbf{n},\sigma}$ is the annihilation operator for a
particle of species $\sigma=1,2$ on the lattice site $\mathbf{n}$. The
$s$-wave interaction is
\begin{equation}
V_{s} = U_s \sum_{\bf n} c^\dagger_{{\bf n},1}c^\dagger_{{\bf n},2} c_{{\bf n},2}c_{{\bf n}, 1 } ~,
\end{equation}
and the $p$-wave interaction is
\begin{eqnarray}
    V_p = U_p\sum_{\mathbf{n}}
\left[
B_x^\dagger(\mathbf{n})B_x(\mathbf{n})
+B_y^\dagger(\mathbf{n})B_y(\mathbf{n})
\right]~,
\end{eqnarray}
with
\begin{eqnarray}
  B_x(\mathbf{n})&=& c_{\mathbf{n},1}c_{\mathbf{n}+\hat x,2}
  -c_{\mathbf{n}+\hat x,1}c_{\mathbf{n},2}~,
\\
B_y(\mathbf{n})&=& c_{\mathbf{n},1}c_{\mathbf{n}+\hat y,2}
              -c_{\mathbf{n}+\hat y,1}c_{\mathbf{n},2}~.
\end{eqnarray}
The operators $B_x$ and $B_y$ are antisymmetric pair operators. They
are odd under particle exchange and therefore generate the
square-lattice analogue of a $p$-wave interaction. The single-particle
dispersion relation for the free Hamiltonian is
\begin{equation}
\epsilon(\qvec)=\frac{1}{m a^2}
\left[(1-\cos q_x)+(1-\cos q_y)\right]~,
\end{equation}
with dimensionless lattice momenta
\begin{equation}
q_i=\frac{2\pi n_i}{N}, \qquad n_i=0,1,\ldots,N-1, \qquad i=x,y.
\end{equation}
For zero total momentum, the two-body relative energy is
\begin{equation}
  E_{\rm rel}(\qvec)=2\epsilon(\qvec)~.
\end{equation}
At low energies this reduces to the continuum expression
\begin{equation}
E \approx \frac{k^2}{m}~.
\end{equation}
The coupling constants $U_s$ and $U_p$ can be fixed by requiring them
to give a specific two-body binding energy. Here, we intentionally
avoid talking about the renormalization: We will treat a change in the
lattice spacing as a modification of the Hamiltonian and thereby
increase the size of the set of {\it models} that are considered.

We write the finite volume, lattice scattering amplitude as
\begin{align}
    T^{(s)}_L(E) = \frac{1}{U_s^{-1}- \Pi_L(E)}~,
\end{align}
where $\Pi_L(E)$ is the finite volume loop function
\begin{align}
    \Pi_L(E) = \frac{1}{N^2}\sum_{\bf n}
    \frac{1}{E-\frac{2}{m a^2}(2-\cos(q_x)-\cos(q_y))}~,
\end{align}
and in the finite volume $q_{x/y} = \frac{2\pi n_{x/y}}{N}$. At the
bound-state pole, $T_L(E)$ diverges and the corresponding pole
condition fixes $U_s$. By requiring that this equation has a pole at
energy $E$ the coupling constant
$U_s$ can be {\it renormalized}.

In the infinite-volume limit the sum becomes an integral over the
Brillouin zone. The corresponding infinite-volume binding energy
$B_\infty$ can then be used to eliminate the bare coupling $U_s$.  In the
infinite volume limit, the two-body loop function is replaced by
\begin{equation}
\Pi_\infty(E)=
\int_{-\pi}^{\pi}\frac{dq_x}{2\pi}
\int_{-\pi}^{\pi}\frac{dq_y}{2\pi}
\frac{1}{
E-\frac{2}{m a^2}\left[2-\cos q_x-\cos q_y\right]
}~.
\end{equation}

At low energies, the loop function develops the universal
two-dimensional logarithmic
structure~\cite{Adhikari:1986,Chadan:1998kq}, and the amplitude
reduces to
\begin{equation}
T_\infty^{(s)}(E)\propto
\frac{1}{\ln(B_\infty/E)-i\pi}~,
\end{equation}
up to normalization conventions. In the continuum limit the pure
$s$-wave differential cross section is isotropic. On the lattice, the
reduced rotational symmetry of the square lattice introduces small
anisotropies of order $(ka)^2$. These lattice artifacts are partially
suppressed by the angular binning used in the construction of our {\it
  detectors} (see Sec.~\ref{sec:observables}) and are small in the
momentum range considered in this work.

To connect this lattice amplitude to the continuum cross section, we
extract the normalization of the amplitude by evaluating the
imaginary part of the loop function $\Pi_\infty^{(s)}$ and use its
inverse as the overall normalization that relates the continuum
amplitude to the lattice amplitude. We find
\begin{equation}
t_0(E) = 
\frac{m a^2}{4} T_\infty^{(s)}(E),
\end{equation}
up to lattice corrections of order $(k a)^2$ that arise from the
energy dependence of the imaginary part of the loop function. We can then use the results derived by
Adhikari~\cite{Adhikari:1986} to relate this amplitude to differential
and total cross section so that
\begin{equation}
  \frac{d\sigma}{d\theta}=\frac{2\pi}{k}|t_0(E)|^2~.
\end{equation}

\paragraph*{\bf P-wave differential cross section -}
When the $p$-wave interaction is included, the differential cross
section acquires angular dependence.  For the pure $p$-wave
interaction, we define the infinite volume lattice $p$-wave amplitude
\begin{equation}
  \label{eq:pwave_amplitude}
T_\infty^{(p)}(E)=\frac{1}{U_p^{-1}-\Pi_\infty^{(p)}(E+i0)}~,
\end{equation}
with the loop function
\begin{equation}
\label{eq:pwave_loop}
  \Pi^{(p)}_\mu(E)=
\int_{-\pi}^{\pi}\frac{dq_x}{2\pi}
\int_{-\pi}^{\pi}\frac{dq_y}{2\pi}
\frac{|f_\mu(\qvec)|^2}{E-E_{\rm rel}(\qvec)+i \epsilon}~.
\end{equation}
The imaginary part of the loop function provides the relative
normalization between the expression in Eq.~\eqref{eq:pwave_amplitude}
and the partial wave projected $p$-wave amplitude $t_1$ defined in
Ref.~\cite{Adhikari:1986}. Using this, we obtain
\begin{equation}
t_1(E)=\frac{m a^4 k^2}{2}T_\infty^{(p)}(E)~,
\end{equation}
so that
\begin{align}
t_1(E)&=-\frac{1}{\cot\delta_1(E)-i}~.
\end{align}
The physical differential cross section is
\begin{equation}
\frac{d\sigma^{(p)}}{d\theta}
=
\frac{2}{\pi k}
\left|
\cos(\theta)\,t_1(E)
\right|^2~,
\end{equation}
where $\theta$ denotes the angle between incoming and outgoing momentum.

\paragraph*{Mixed $s{+}p$ differential cross section.}
Since we are able to relate our lattice amplitudes to Adhikari's
partial wave amplitudes, it is also easy to calculate the infinite
volume differential cross section for the case of the mixed $s$- and
$p$-wave Hamiltonian. We use
\begin{equation}
\label{eq:dsigma_mixed}
  \frac{d\sigma^{(s+p)}}{d\theta}
=
\frac{2}{\pi k}
\left|
t_0(E)+2\cos(\theta)\,t_1(E)
\right|^2~.
\end{equation}

\section{Initial states and real-time evolution}
\label{sec:time-evolution}
We initialize the system as a product of one-body wave packets,
\begin{equation}
\Psi(\rvec_1,\rvec_2,0)=\phi_1(\rvec_1)\phi_2(\rvec_2),
\end{equation}
with particle~1 and particle~2 placed symmetrically about the box
center and given opposite mean momenta. In the runs used for
the datasets, the initial separation is held fixed in physical units
while the box size is varied. This is crucial for meaningful
finite-volume comparisons: one wants to enlarge the box around the
same physical scattering setup, rather than change the initial state
together with the volume.

The one-body packets are either Gaussian, labeled later in datasets with
$\mathcal{F}_0 =0$, or super-Gaussian,
$\mathcal{F}=1$. For a packet centered at~$\rvec_0$ with mean
momentum~$\kvec_0$, the periodic Gaussian envelope is
\begin{equation}
\phi_G(\rvec)=\mathcal N_G
\exp\!\left[-\frac{|\rvec-\rvec_0|_{\rm min}^2}{4\sigma^2}
+i\kvec_0\cdot\rvec\right],
\end{equation}
where $|\rvec-\rvec_0|_{\rm min}$ denotes the minimum-image wrapped
distance. The super-Gaussian generalization is
\begin{equation}
\phi_{SG}(\rvec)=\mathcal N_{SG}
\exp\!\left[-\left(\frac{|\rvec-\rvec_0|_{\rm min}}{\sigma}\right)^n
+i\kvec_0\cdot\rvec\right],
\end{equation}
with power $n\ge 2$. The latter are useful because they allow us to
vary the packet shape and momentum spread while keeping the
Hamiltonian fixed. In both cases the normalization constants $\mathcal N_G$ and $\mathcal N_{SG}$ are
chosen so that the one-body lattice states satisfy
\begin{equation}
\sum_{\rvec} |\phi(\rvec)|^2 = 1~,
\end{equation}
and hence the two-body initial state is normalized to unity.

The real-time evolution is generated by the lattice Hamiltonian,
\begin{equation}
\Psi(t)=e^{-iHt}\Psi(0),
\end{equation}
and implemented numerically by a second-order split-operator scheme on
the sparse two-body lattice Hamiltonian. Writing $H=T+V$ with the
kinetic part $T$ stored as a sparse matrix and the interaction in a
diagonal vector $V$, each time step is approximated as
\begin{equation}
e^{-iH\Delta t}\approx
e^{-iV\Delta t/2}\,e^{-iT\Delta t}\,e^{-iV\Delta t/2}~.
\end{equation}
This time evolution is easily implemented using matrix exponential
routines. We emphasize that we do not propagate the wave
packet through exact diagonalization.  Instead, we carry out the
evolution directly on the lattice by repeated split-operator
updates. At each stored time step we then record the wedge-resolved
detector probabilities defined in Sec.~\ref{sec:observables}.
\begin{figure}[t]
    \centerline{\includegraphics[width=0.95\columnwidth]{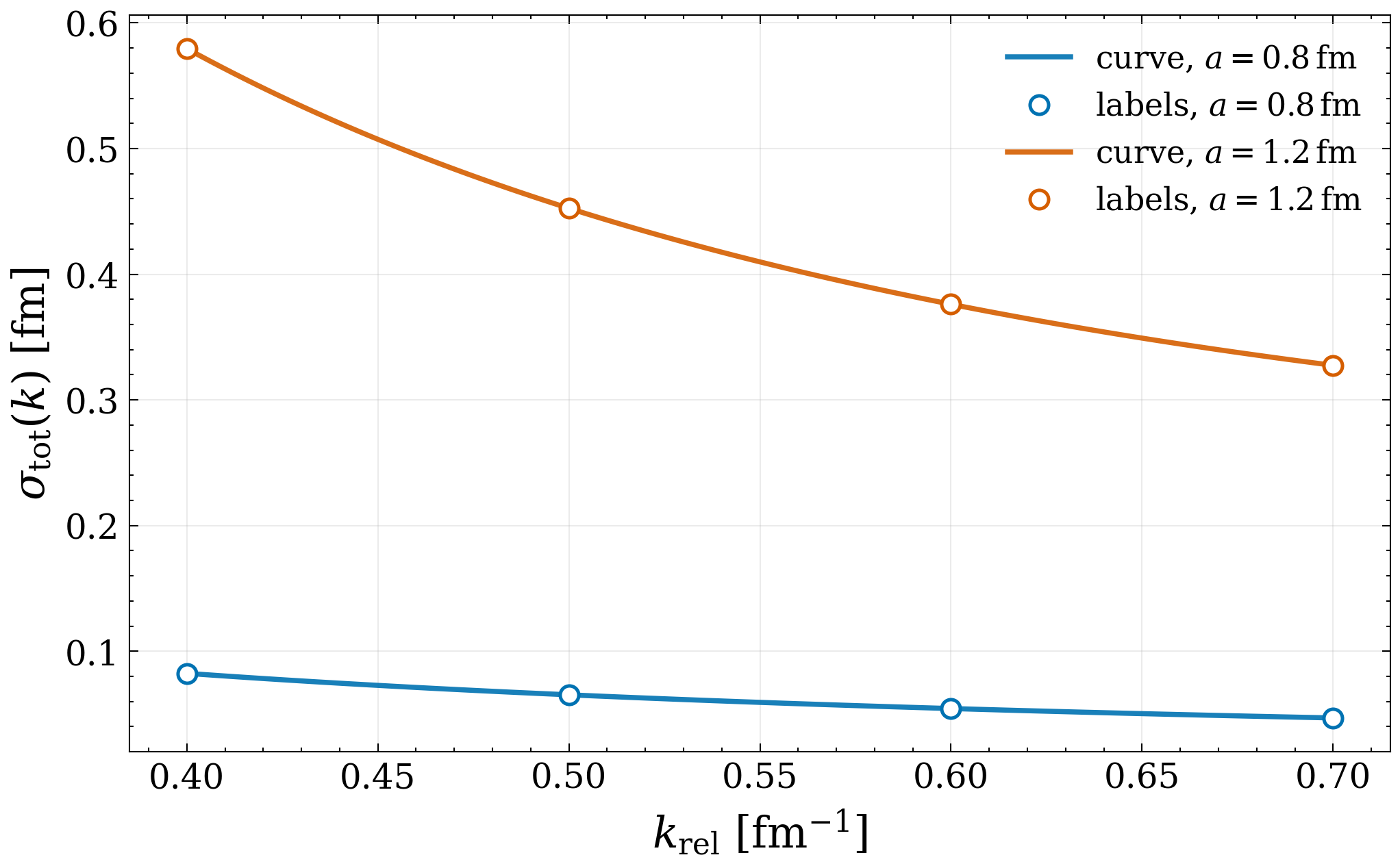}}
    \caption{Infinite-volume pure $s$-wave total cross section as a
      function of the incoming relative momentum for a representative
      $U_s=-20~{\rm MeV}$ coupling. For each lattice spacing $a$, the
      solid curve shows the infinite-volume lattice result obtained
      from the Brillouin-zone integral. The open circles show the
      values that are part of the training dataset.}
    \label{fig:iv-target}
\end{figure}

\section{Detector wedges and finite-volume observables}
\label{sec:observables}

The {\it detected}  observables in this work are defined in the relative
coordinate. In continuum scattering, the differential cross section
determines the angular distribution of scattered flux at large
distances from the interaction region. Our present approach is
designed to capture the same information on a finite lattice: we
partition an annulus in the relative coordinate into angular sectors
and record the probability flowing into each sector as a function of
time. We refer to these sectors as detector wedges because they are
wedge-shaped regions of the annulus.

Because the lattice is periodic, the relative distance is defined
as the shortest distance among all periodic copies. A detector
wedge $D_\alpha$ is the subset of configuration-space points for which
the relative vector obeys
\begin{align}
  \label{eq:wedge_def}
  \nonumber
R_1 &< |\rvec_1-\rvec_2| < R_2, \\
\theta_\alpha &< \arg(\rvec_1-\rvec_2) < \theta_\alpha+\Delta\theta,
\end{align}
with $\Delta\theta=2\pi/N_\theta$ for a uniform $N_\theta$-bin
partition of the annulus. In the production datasets used here we take
$N_\theta=12$. The inner cutoff $R_1$ excludes the immediate
interaction region, while $R_2$ selects an annular shell in relative
space. In the production datasets used here we take
$R_1=1.6~{\rm fm}$ and $R_2=3.5~{\rm fm}$, and both radii are held
fixed in physical units across all box sizes.

The probability associated with wedge $\alpha$ is
\begin{equation}
P_\alpha(t)=
\sum_{(\rvec_1,\rvec_2)\in D_\alpha}
|\Psi(\rvec_1,\rvec_2,t)|^2.
\end{equation}
We also compute the same quantity for the noninteracting problem,
$P_\alpha^0(t)$, using the same initial state and the same detector
geometry. Their difference,
\begin{equation}
\Delta P_\alpha(t)=P_\alpha^0(t)-P_\alpha(t),
\end{equation}
isolates the redistribution caused by the interaction. Since the
absolute detector probabilities can vary with box size, packet width,
and total probability flowing through the annulus, it is also useful
to study the normalized interacting wedge fractions
\begin{equation}
\widetilde P_\alpha(t)=
\frac{P_\alpha(t)}{\sum_\beta P_\beta(t)}.
\label{eq:normalized-wedge-fraction}
\end{equation}
In practice, we find that these normalized wedge fractions converge
faster with box size than the raw $P_\alpha(t)$ values, which makes
them especially useful for finite-volume comparisons and for
data-driven inference. Even for an isotropic infinite-volume $s$-wave
target, the finite-time wedge signal need not be isotropic: the
incoming packets select a beam axis, the detector samples a finite
annulus, and the observable includes both incident and scattered
contributions inside a periodic box.

The physical motivation for using wedge observables as predictors of
the infinite-volume cross section is as follows. When the wave packets
collide, the interaction redistributes probability among the angular
wedges relative to the noninteracting case. A stronger interaction
produces a larger overall redistribution, while the angular pattern of
$\Delta P_\alpha(t)$ reflects the angular dependence of the underlying
scattering amplitude. In the infinite volume limit and at
asymptotically late times, this angular redistribution would converge
to the differential cross section up to kinematic prefactors. In a
finite periodic box, however, the relationship is complicated by
boundary effects, including the return of scattered wave packets
through the periodic boundaries, and by the finite momentum spread of
the initial state.  Rather than attempting to extract the cross
section from an approximate asymptotic formula, we adopt a data-driven
approach: the CNN learns the mapping from the full wedge-time signal
$\{\Delta P_\alpha(t),\widetilde P_\alpha(t)\}$ to the infinite-volume
cross section, using supervised labels computed within the same model.

The coarse angular binning also plays a useful role in controlling
lattice artifacts. On the square lattice, the reduced rotational
symmetry introduces anisotropies. Integrating the probability density
over $30^\circ$ wedges partially averages over these artifacts.

\section{Benchmarks and machine-learning results}
\label{sec:results}

\paragraph*{\bf Validation  - }
We verified that the periodic pole equation reproduces the binding
energies from direct diagonalization of the Hamiltonian for all tested
box sizes. This indicates that the infinite-volume contact labels that
we compute are quantitatively consistent with the same lattice
Hamiltonian used to generate the finite-volume time-dependent data.

For the pure $s$-wave problem, we have used continuum calculations
with a separable interaction and lattice calculations, both
renormalized to give the same binding energy, to verify that our
relation between lattice and continuum amplitude is correct.

The pure $s$-wave model gives an infinite-volume cross section whose
magnitude depends on the incoming relative momentum. In
Fig.~\ref{fig:iv-target} we show the $s$-wave model cross section as a
function of the relative momentum $k$ for a representative coupling
$U_s=-20$~MeV and for the two lattice spacings that will be used in
the $s$-wave dataset. The cross section varies strongly with momentum,
which is what is needed for a meaningful supervised-learning problem.

We also studied the convergence of our wedge detector signals: We
generated finite-volume two-body datasets on $12\times 12$,
$14\times 14$, $16\times 16$, and $18\times 18$ lattices and varied
the incoming momentum, packet width, and packet family while keeping
the detector geometry fixed. After removing redundant head-on
configurations related by particle-label exchange, each box-size
campaign contains 63 samples. The campaigns for the three different
box sizes provide therefore a combined dataset of $N_{\rm samp}=189$
samples covering $k=0.4,0.6,0.8~{\rm fm}^{-1}$, packet widths
$\sigma=0.8,1.0~{\rm fm}$, Gaussian and super-Gaussian envelopes, and
box sizes $12\times 12$, $14\times 14$, and $16\times 16$.

\begin{figure}[t]
    \centerline{\includegraphics[width=0.95\columnwidth]{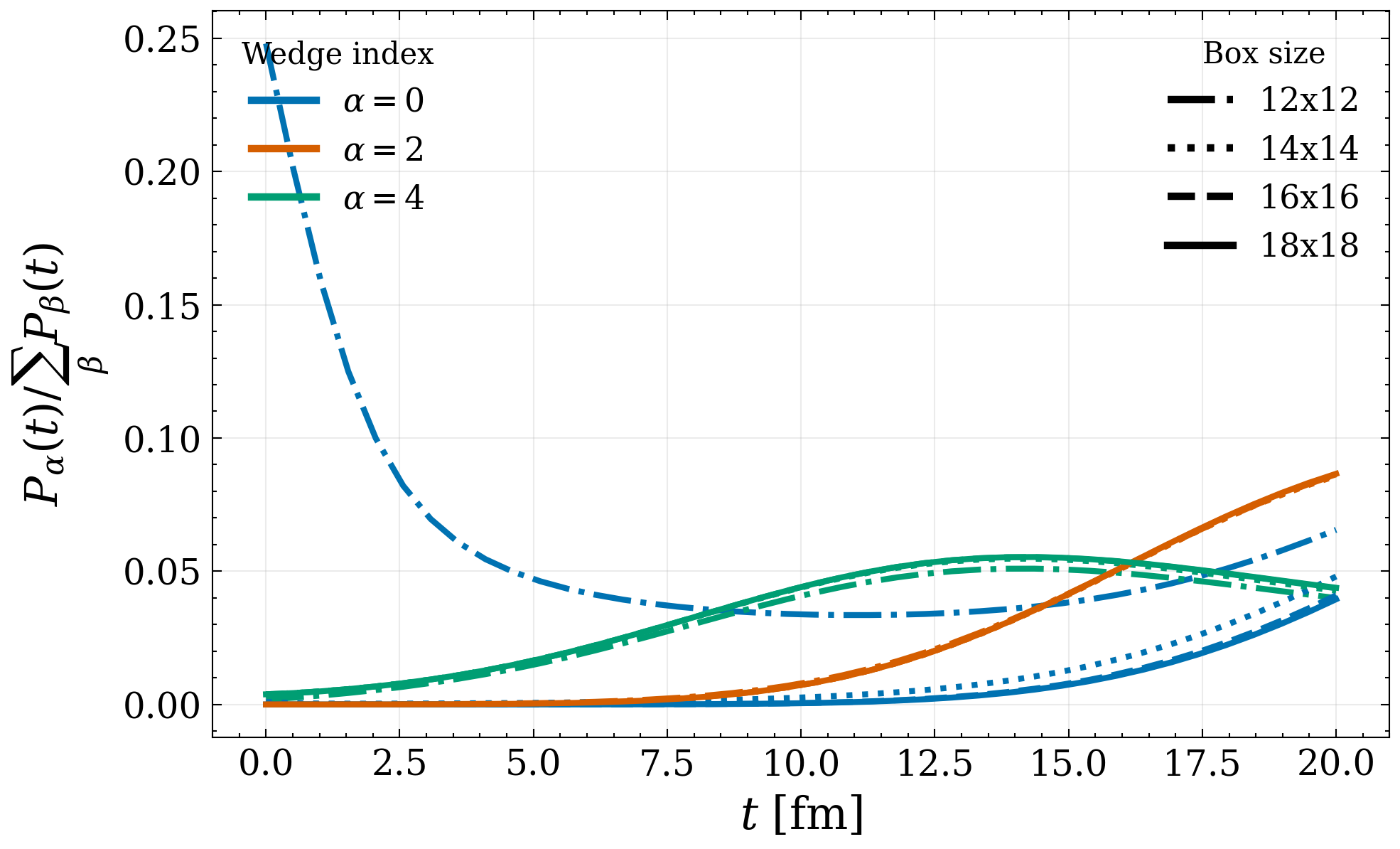}}
    \caption{Normalized wedge fractions for fixed physical initial
      separation, fixed detector radii and across different box volumes.
      The results shown were obtained from the same representative head-on event with
      $U_s=-20~{\rm MeV}$, $a=1.0~{\rm fm}$,
      $k=0.4~{\rm fm}^{-1}$, $\sigma_1=\sigma_2=0.8~{\rm fm}$,
      and Gaussian wave packets for both particles.
    }\label{fig:wedge-convergence}
\end{figure}

Figure~\ref{fig:wedge-convergence} shows the convergence behavior of
the time dependence of the normalized wedge fractions for a specific
choice of wave packet parameters across different box volumes. The
wedge signals converge with increasing box volume which indicates that
at sufficiently large box sizes we are in a meaningful finite volume
regime. We furthermore do not see clear oscillatory behavior in the
time window considered, which would indicate significant amounts of
probability crossing the boundaries.

\paragraph*{\bf Data construction -} The next component of our
approach is the construction of the training samples supplied to the
CNN. Each sample contains the wedge-resolved time series, the
non-interacting reference, diagnostic observables (norms, one-body
densities, contact probabilities, and mean relative separations), and
the attached infinite-volume cross sections. We then assemble
these samples into training, validation, and testing sets.  The
resulting input tensor has shape
$N_{\rm samp}\times N_{\rm ch}\times N_t\times N_\theta$, where
$N_{\rm samp}$ is the number of samples, $N_{\rm ch}$ is the number of
input channels, $N_t$ is the number of stored time steps, and
$N_\theta$ is the number of angular wedges.
In all datasets that will be discussed below we used lattice spacings
$a = 0.8, 1.2$~fm, widths $\sigma = 0.7,0.9, 1.1$~fm, box sizes
$N=8,10$, relative momenta $k = 0.4,0.5,0.6,0.7$~fm$^{-1}$, and
Gaussian and super-Gaussian wave packet families with super-Gaussian
powers 2 and 4.

The neural networks used in this work are convolutional models acting
on the wedge-time tensor of shape $(N_{\rm ch},N_t,N_\theta)$. Their
architecture is summarized in Tbl.~\ref{tab:cnn-architecture}. In the
experiment-informed and fully conditioned variants, the additional
input parameters (for example the incoming momentum, packet width, box size,
and interaction parameters) enter only at the final prediction stage, after the
convolutional processing of the wedge-time tensor. For the mixed
$s{+}p$ benchmark, the network uses two output branches, one for the
total cross section and one for the normalized angular shape, whose
predictions are combined to reconstruct the full binned differential
cross section.

All CNNs are trained with the AdamW optimizer and mean-squared-error
loss. For the pure $s$- and pure $p$-wave benchmarks, the models
considered here use batch size 16, weight decay $10^{-4}$, and
learning rates $3\times 10^{-3}$ for $(C,H)=(8,32)$ and
$2\times 10^{-3}$ for $(C,H)=(12,48)$. These models are trained for up
to 200 epochs, with early stopping based on the validation loss if no
improvement is seen for 20 epochs.  For the mixed $s{+}p$ benchmark,
the loss is the sum of the mean-squared errors for the
total-cross-section and normalized-shape outputs, with equal
weights. The mixed models use batch size 16, weight decay $10^{-4}$ (a
penalty on large network weights), learning rate $2\times 10^{-3}$,
and the fixed architecture $(C,H)=(12,48)$, and are trained for 30
epochs without early stopping in order to keep the comparison between
the different mixed-benchmark variants uniform.

\begin{table}[t]
\begin{ruledtabular}
\begin{tabular}{p{0.34\columnwidth}p{0.58\columnwidth}}
Component & Architecture \\
\hline
Input & wedge-time tensor $(N_{\rm ch},N_t,N_\theta)$ with $N_{\rm ch}=2$ \\
Convolution 1 & 2D convolution from 2 input channels to $C$ feature channels, kernel $3\times 3$, padding 1, followed by GELU activation \\
Convolution 2 & 2D convolution from $C$ channels to $C$ channels, kernel $3\times 3$, padding 1, followed by GELU activation \\
Pooling & max pooling with kernel size 2 \\
Convolution 3 & 2D convolution from $C$ channels to $2C$ channels, kernel $3\times 3$, padding 1, followed by GELU activation \\
Feature compression & adaptive average pooling to a $3\times 3$ feature map, followed by flattening \\
Final prediction stage & one fully connected hidden layer of width $H$, followed by GELU activation; in the experiment-informed and fully conditioned models, the additional input parameters enter at this stage \\
Output: pure $s$ and pure $p$ & single output branch for the reference cross section \\
Output: mixed $s{+}p$ & two output branches: one for the total cross section and one for the normalized angular shape \\
\end{tabular}
\end{ruledtabular}
\caption{Compact summary of the CNN architectures used in this work.
For the pure $s$-wave benchmark, the reported models use $(C,H)=(8,32)$
for the dynamics-only variant and $(12,48)$ for the experiment-informed
and fully conditioned variants. For the pure $p$-wave benchmark, the
dynamics-only model uses $(8,32)$ and the experiment-informed and fully
conditioned models use $(12,48)$. For the mixed $s{+}p$ benchmark, all
reported models with two output branches use the fixed choice $(C,H)=(12,48)$. Here
$C$ denotes the number of convolutional feature channels and $H$ the
width of the hidden fully connected layer.}
\label{tab:cnn-architecture}
\end{table}

\paragraph*{\bf Evaluation metrics and shape reconstruction}
We will quantify the quality of the neural network predictions using
standard tools from statistics. Specifically, we will compare
predictions for held out data with the data itself. The first metric
that we will use, the so-called {\it coefficient of determination} or
$R^2$, measures how well a model captures the variation of the reference
data. For a data set $y_n$ and model predictions $\hat y_n$, it is
defined as
\begin{align}
  \label{eq:rsquared}
\nonumber
  R^2
&=
1-\frac{\sum_n\left(\hat y_n-y_n\right)^2}
{\sum_n\left(y_n-\bar y\right)^2}~,
\\
\bar y&=\frac{1}{N_{\rm eval}}\sum_n y_n~,
\end{align}
where $N_{\rm eval}$ denotes the number of evaluated model samples in
the corresponding test set. An $R^2$ value close to one indicates a
good description of the data. As a second metric, we will also give is
the mean relative error (MRE) of our predictions relative to the held
out data. It is defined as
\begin{equation}
  \label{eq:mre}
{\rm MRE}
=
\frac{1}{N_{\rm eval}}\sum_n
\frac{|\hat y_n-y_n|}{|y_n|}~.
\end{equation}
For angle-dependent observables it is useful to separate the total
magnitude from the normalized angular profile. If the binned
differential cross section is denoted by
$(d\sigma/d\theta)_\alpha$, we define the corresponding normalized
angular shape by
\begin{equation}
s_\alpha =
\frac{(d\sigma/d\theta)_\alpha}
{\sum_{\beta=1}^{N_\theta}(d\sigma/d\theta)_\beta},
\qquad
\sum_{\alpha=1}^{N_\theta} s_\alpha = 1~.
\end{equation}
When we quote the angular-shape coefficient of determination, we mean
\begin{align}
\nonumber
  R^2_{\rm shape}
&=
1-\frac{\sum_{n=1}^{N_{\rm eval}}\sum_{\alpha=1}^{N_\theta}(\hat s_{n\alpha}-s_{n\alpha})^2}
{\sum_{n=1}^{N_{\rm eval}}\sum_{\alpha=1}^{N_\theta}(s_{n\alpha}-\bar s)^2}~,\\
\bar s&=\frac{1}{N_{\rm eval}N_\theta}\sum_{n=1}^{N_{\rm eval}}\sum_{\alpha=1}^{N_\theta}s_{n\alpha}.
\end{align}
Thus $R^2_{\rm shape}=1$ corresponds to a perfect prediction of the
normalized angular profile alone, independent of the overall scale.

\paragraph*{\bf Pure $s$-wave scattering} To test whether the
finite-volume time-dependent signal can predict an unseen interaction
strength, we constructed a dataset in which the coupling strength
$U_s$, lattice spacing $a$, box size $L=Na$, packet width, and
incoming momentum are varied together. This dataset spans
$U_s=-10,-15,-20,-25,-30,-35~{\rm MeV}$ while the other parameters
were varied as described above.

From this dataset we then define a held-out coupling benchmark by
choosing $U_s=-20~{\rm MeV}$ as the held-out value. The CNN is trained
on all samples at $U_s=-10,-15,-25,-30,-35~{\rm MeV}$ and evaluated on
the previously unseen coupling $U_s=-20~{\rm MeV}$. The input is the
wedge-time tensor built from $\Delta P_\alpha(t)$ and the normalized
wedge fractions. The model used below is a small CNN acting directly
on the $(N_{\rm ch},N_t,N_\theta)$ tensor.

\begin{figure}[t]
    \centerline{\includegraphics[width=0.95\columnwidth]{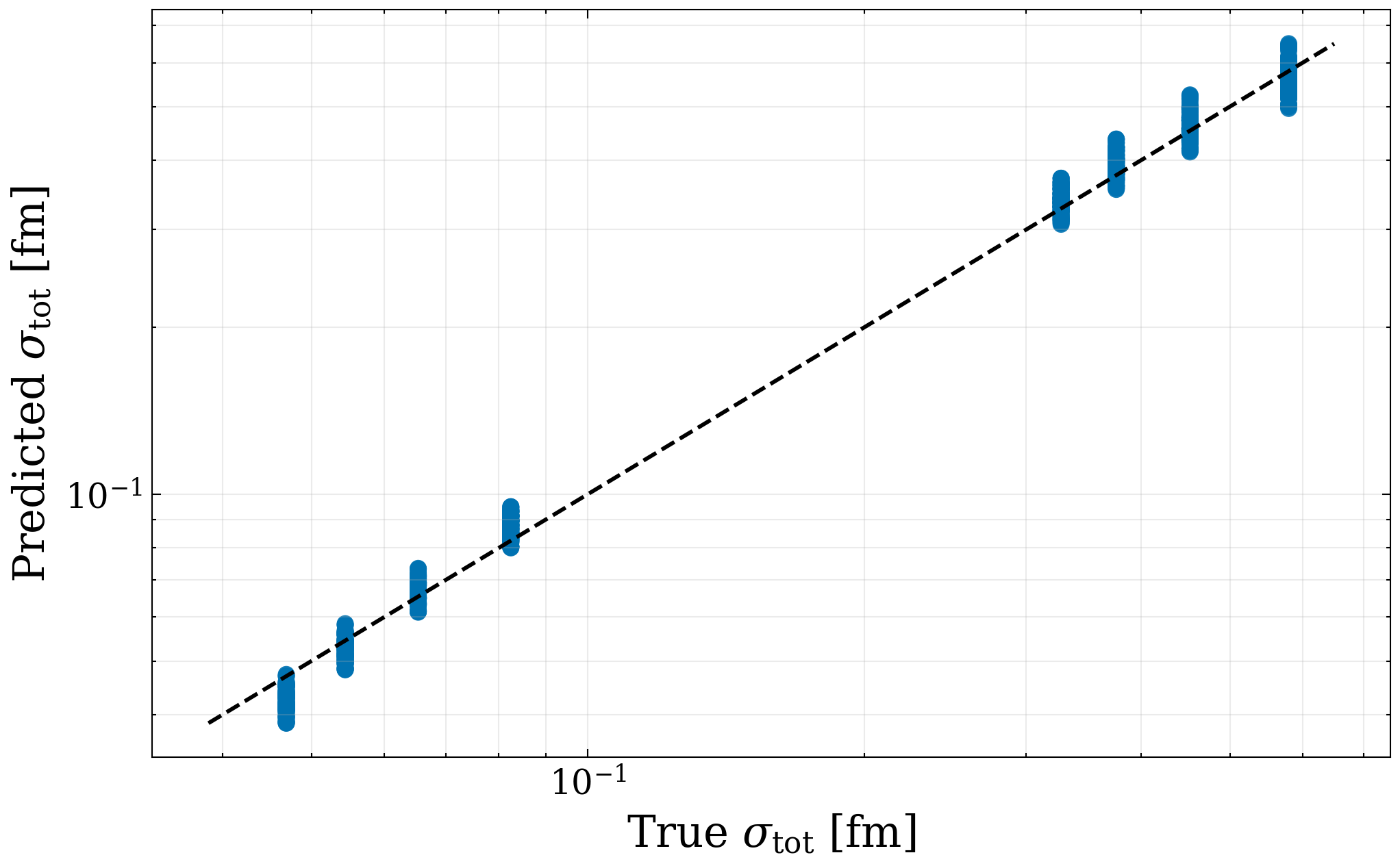}}
    \centerline{\includegraphics[width=0.95\columnwidth]{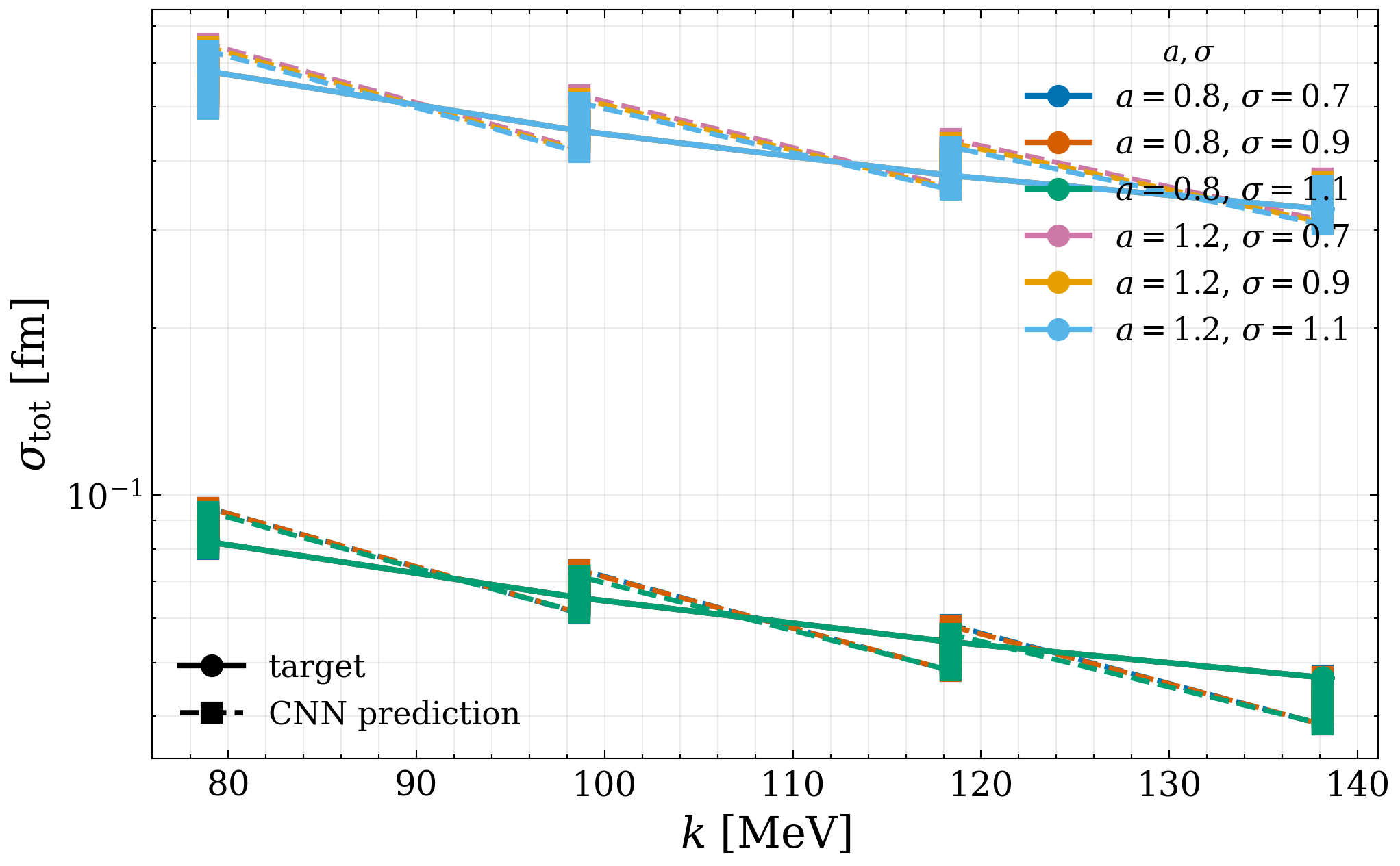}}
    \caption{Comparison of CNN prediction and held out cross section
      for the pure $s$-wave model with $U_s = -20$~MeV. Top:
      Comparison of the predicted and reference cross sections for all
      held-out samples; the dashed diagonal shows the line that
      indicates perfect agreement. Bottom: reference and predicted cross
      sections as a function of the incoming relative momentum $k$ for
      the held-out samples. Colors denote the lattice spacing and
      packet width combination $(a,\sigma)$, solid circles mark the
      reference values and dashed squares mark the CNN predictions.}
    \label{fig:ai-parity}
\end{figure}

We also tested simpler comparison models, including fits to the same
wedge-time data and a parameter-only baseline that used only known
parameters of the calculation without detector information. None of
these matched the CNN performance.  Figure~\ref{fig:ai-parity} shows
the corresponding prediction-versus-reference plot and energy slices for
this held-out-$U_s$ test. The prediction-versus-reference panel (top)
shows that the CNN predictions lie close to the diagonal, while bottom
panel shows that the CNN reproduces the momentum dependence of the
held-out cross section across the different lattice spacings and
packet widths.  The six-coupling dataset shows that the observable is
learnable once the coupling coverage is sufficiently dense. We train
three different models: (i) a model that receives only detector
information that we will call {\it dynamics only}, (ii) a model that
receives the initial state and detector information, and (iii) a model
that receives all information including the Hamiltonian parameters
which we will call {\it fully conditioned}.  We furthermore compare
against the parameter-only baseline described above. Details regarding
the input to these models is also provided in
Tbl.~\ref{tab:spwave-inputs}. We find that the fully conditioned model
performs best, reaching $R^2=0.989$ with mean relative error
$5.74\times 10^{-2}$. The experiment-informed and dynamics-only
variants degrade to $R^2=0.973$ and $0.952$, with mean relative errors
$7.77\times 10^{-2}$ and $9.99\times 10^{-2}$, respectively.  The
parameter-only baseline reaches $R^2=0.909$ with mean relative error
$1.95\times 10^{-1}$. The corresponding benchmark metrics are
collected in Tbl.~\ref{tab:benchmark-metrics}.

The infinite volume cross section is not angle dependent and
significantly easier to {\it learn} than an angle dependent
quantity. It is however, remarkable, that a rather limited amount of
data that does not have any information about the Hamiltonian is able
to predict a scattering observable to relatively high accuracy.

\begin{table*}[t]
\label{tab:spwave-inputs}
\begin{ruledtabular}
\begin{tabular}{p{0.24\textwidth}p{0.66\textwidth}}
  Model & Input \\
  \hline
  Dynamics only & $\{\Delta P_\alpha(t),\, \widetilde P_\alpha(t)\}$ \\
  Experiment informed & $\{\Delta P_\alpha(t),\, \widetilde P_\alpha(t),\, k_{ix}, \sigma_i, {\cal F}_i,\, n_i,\, d,\, N \}$, where ${\cal F}_i=0, 1$  \\
  Fully conditioned & $\{\Delta P_\alpha(t),\, \widetilde P_\alpha(t),\, k_{ix}, \sigma_i, {\cal F}_i,\, n_i,\, d,\, N,\, a,\, U_s,\, U_p\}$. \\
  \hline
  Baseline (parameter only) & $\{k_{ix}, \sigma_i, {\cal F}_i\, n_i, d,\, N,\, a,\, U_s,\, U_p\}$\\
\end{tabular}
\end{ruledtabular}
\caption{Input sets used in the benchmark families. For the particle index we have $i=1,2$. The three CNN
  variants all receive $\Delta P_\alpha(t)$ and
  $\widetilde P_\alpha(t)$ and differ otherwise only in the additional
  input parameters they receive. The parameter-only baseline uses the same
  input parameters without the wedge-time tensor.}
\end{table*}

\paragraph*{\bf Pure $p$-wave scattering -} The next step is to test
whether the same strategy can produce predictions for the
angle-dependent $p$-wave differential cross section. To test this, we
generate a dataset with the s-wave part of the Hamiltonian set to
zero.  This dataset includes the $p$-wave coupling values
$U_p=-90,-100,-105,-110,-115,-120~{\rm MeV}$.
Our resulting dataset contains 4320 samples. We hold out the data with
$U_p=-110~{\rm MeV}$ for testing and use the remaining couplings for
training and validation. We emphasize that this includes all
combinations of box sizes, lattice spacings and all other
parameters. Specifically, we use 2880 training samples, 720 validation
samples, and 720 held-out test samples.

The pure $p$-wave CNN receives the 12 infinite-volume differential cross
sections corresponding to the wedges used in the dataset. For the
dynamics-only model, the stronger-coupling held-out case
$U_p=-110~{\rm MeV}$ fails, with $R^2=-1.763$ and mean relative error
$1.00$. By contrast, the same held-out benchmark is
accurately recovered once additional input parameters are supplied: the
experiment-informed model reaches $R^2=0.999998$ with mean relative
error $1.16\times 10^{-3}$, and the fully conditioned model reaches
$R^2=0.999983$ with mean relative error $3.53\times 10^{-3}$. The
corresponding benchmark metrics are collected in
Tbl.~\ref{tab:benchmark-metrics}.

\paragraph*{\bf Mixed $s{+}p$ scattering -}
Finally, we consider the case in which $s$- and $p$-wave
interaction are mixed. We therefore extend the dataset to include
combinations of non-zero $U_s$ and $U_p$. The infinite volume
differential cross sections are then calculated numerically using
Eq.~\eqref{eq:dsigma_mixed}.

The dataset includes the coupling constant values
$U_s=-10,-15,-20,-25,-30,-35~{\rm MeV}$ and
$U_p=-90,-100,-105,-110,-115,-120~{\rm MeV}$.
The resulting supervised dataset contains 25920 samples.
We use one pair of coupling constants for testing and use the
remaining samples for training and validation. For this problem it
is more natural to learn the total magnitude and the normalized
angular shape separately and then reconstruct the full 12-bin
differential cross section. We therefore use a CNN with two output branches, one
for the total cross section and one for the normalized
12-bin angular distribution. The input tensor has shape
$(N_{\rm ch},N_t,N_\theta)$, with the two feature channels given by
$\Delta P_\alpha(t)$ and $\widetilde P_\alpha(t)$. The network
predicts the total cross section and the normalized shape separately,
and the full binned prediction is reconstructed as
\begin{equation}
\left(\frac{d\hat\sigma}{d\theta}\right)_\alpha
=
\hat \sigma_{\rm tot}\,\hat s_\alpha.
\end{equation}
The same pooled convention is used for the pure $p$ benchmark, where
the quoted $R^2$ is computed from the full 12-bin differential cross
sections. We then test three input regimes for the held-out mixed
Hamiltonian pair $(U_s,U_p)=(-20,-110)~{\rm MeV}$.
Specifically, we use 20160 training samples, 5040 validation samples,
and 720 held-out test samples.
For nucleon-mass particles this corresponds to relative kinetic
energies of roughly $6.6$--$20.3~{\rm MeV}$.
The held-out point lies in the interior of the scanned coupling grid
and therefore represents interpolation within the present Hamiltonian
family rather than edge extrapolation.

On the held-out test set, the fully conditioned model performs best,
reaching $R^2=0.99988$ with mean relative error
$9.34\times 10^{-3}$ and normalized-shape score
$R^2_{\rm shape}=0.99985$. The experiment-informed model reaches
$R^2=0.99902$, mean relative error $2.06\times 10^{-2}$, and
$R^2_{\rm shape}=0.99886$, while the dynamics-only model gives
$R^2=0.99662$, mean relative error $3.55\times 10^{-2}$, and
$R^2_{\rm shape}=0.99654$. For comparison, the parameter-only baseline
gives $R^2=0.99687$, mean relative error $3.84\times 10^{-2}$, and
$R^2_{\rm shape}=0.99783$.

In all three cases, we use the same infinite-volume 12-bin
differential cross section. The first CNN uses only the finite-volume
wedge-time tensor. The second is experiment-informed: it receives the
same tensor together with the initial state information, {\it i.e.}
the incoming momenta, packet widths, packet families, packet-shape
powers, initial separation, and box size $N$, but not the interaction
strengths $U_s$ and $U_p$. The third network is
fully conditioned and uses all available data including the parameters
$(U_s,U_p)$. We also compare against a parameter-only control, meaning
a model that receives only the preparation and Hamiltonian parameters,
without the wedge-time tensor, in order to test how much predictive
power comes from the finite-volume dynamical signal itself. In this
context packet family labels the envelope type, such as Gaussian or
super-Gaussian, while the packet-shape power is the exponent $n_i$ in
the super-Gaussian profile.

\begin{table*}[t]
\begin{ruledtabular}
\begin{tabular}{p{0.16\textwidth}p{0.18\textwidth}p{0.24\textwidth}ccc}
Benchmark & Model or regime & Reference observable & MRE & $R^2$ & $R^2_{\rm shape}$ \\
\hline
$s$-wave & Dynamics only & $\sigma_{\mathrm{tot}}$ & $9.99\times 10^{-2}$ & $0.9521$ & - \\
$s$-wave & Experiment informed & $\sigma_{\mathrm{tot}}$ & $7.77\times 10^{-2}$ & $0.9726$ & - \\
$s$-wave & Fully conditioned & $\sigma_{\mathrm{tot}}$ & $5.74\times 10^{-2}$ & $0.9886$ & - \\
$s$-wave & Baseline & $\sigma_{\mathrm{tot}}$ & $1.95\times 10^{-1}$ & $0.9093$ & - \\
$p$-wave & Dynamics only & $\textstyle{\left(\frac{d\sigma}{d\theta}\right)_\alpha}$ & $1.00$ & $-1.763$ & - \\
$p$-wave & Experiment informed & $\textstyle{\left(\frac{d\sigma}{d\theta}\right)_\alpha}$ & $1.16\times 10^{-3}$ & $0.999998$ & - \\
  $p$-wave & Fully conditioned & $\textstyle{\left(\frac{d\sigma}{d\theta}\right)_\alpha}$& $3.53\times 10^{-3}$ & $0.999983$ & - \\
$p$-wave & Baseline & $\textstyle{\left(\frac{d\sigma}{d\theta}\right)_\alpha}$ & $3.54\times 10^{-2}$ & $0.99681$ & - \\
$(s{+}p)$-wave & Dynamics only & $\textstyle{\left(\frac{d\sigma}{d\theta}\right)_\alpha}$& $3.55\times 10^{-2}$ & $0.99662$ & $0.99654$ \\
$(s{+}p)$-wave & Experiment informed & $\textstyle{\left(\frac{d\sigma}{d\theta}\right)_\alpha}$ & $2.06\times 10^{-2}$ & $0.99902$ & $0.99886$ \\
$(s{+}p)$-wave & Fully conditioned & $\textstyle{\left(\frac{d\sigma}{d\theta}\right)_\alpha}$& $9.34\times 10^{-3}$ & $0.99988$ & $0.99985$ \\
$(s{+}p)$-wave & Baseline & $\textstyle{\left(\frac{d\sigma}{d\theta}\right)_\alpha}$ & $3.84\times 10^{-2}$ & $0.99687$ & $0.99783$ \\
\end{tabular}
\end{ruledtabular}
  \caption{Benchmark metrics for the pure $s$, pure $p$, and mixed
    $s{+}p$ datasets. Shown are the reference observable, model class, mean
    relative error (MRE), the reference-cross-section score $R^2$, and, for the
    mixed benchmark with two output branches, the normalized-shape score
    $R^2_{\rm shape}$. The parameter-only baseline uses the full
    auxiliary input for each benchmark.}
\label{tab:benchmark-metrics}
\end{table*}

\begin{figure*}[t]
    \centering
    \includegraphics[width=\textwidth]{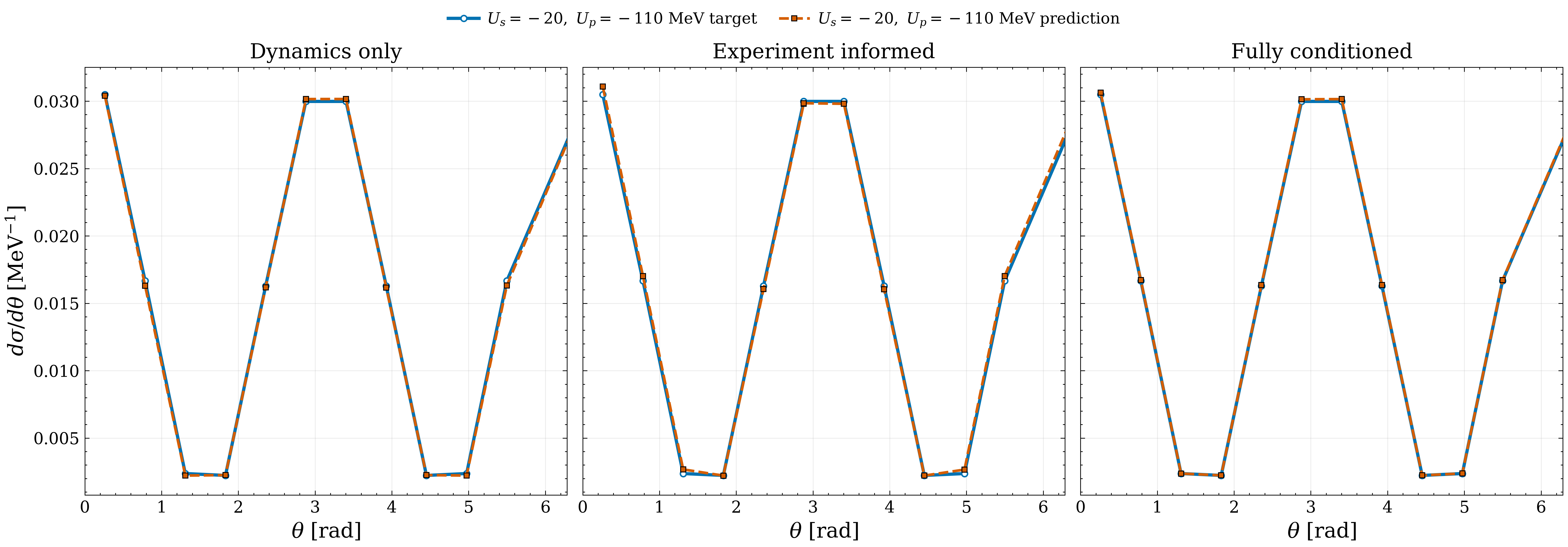}
    \caption{Mixed $(s+p)$ results for the held-out
      Hamiltonian pair $(U_s,U_p)=(-20,-110)~{\rm MeV}$. Left:
      dynamics-only CNN with two output branches trained only on the finite-volume
      wedge-time tensor. Center: experiment-informed CNN with two output branches
      trained on the same dataset together with initial state
      and box information, but without the interaction
      strengths. Right: fully conditioned CNN with two output branches trained on the
      $(s+p)$ dataset, including $(U_s,U_p)$. In each panel the solid
      curves show the reference differential cross sections and dashed
      curves for the CNN predictions. See
      Tbl.~\ref{tab:spwave-inputs} for the complete input lists.}
    \label{fig:spwave-cnn}
\end{figure*}

On the held-out test set the fully conditioned model is best. Removing
the explicit Hamiltonian input and then the remaining preparation
information leads to a controlled degradation in both the recovered total
magnitude and the angular shape. Relative to the dynamics-only model,
the experiment-informed model improves both the total-scale error and
the angular reconstruction, while the parameter-only control provides a
useful baseline for both observables. The corresponding benchmark
metrics are collected in Tbl.~\ref{tab:benchmark-metrics}. Figure~\ref{fig:spwave-cnn}
shows representative held-out samples from the held-out Hamiltonian pair in
every panel for the three wedge-time CNN variants: the dynamics-only
model, the experiment-informed model, and the fully conditioned model.
The parameter-only control is not shown in the figure, but its
performance is included in Tbl.~\ref{tab:benchmark-metrics}.

The comparison between the dynamics-only and parameter-only baseline,
with access to the full Hamiltonian, shows that the finite-volume
time-dependent signal carries predictive information comparable to the
Hamiltonian itself. This is an important outcome of the present
study. On an enlarged Hamiltonian family with a genuinely
angle-dependent observable, the finite-volume wedge-time tensor
predicts both the overall scale and the angular structure of the
infinite-volume differential cross section for previously unseen
Hamiltonians.  The pure $p$-wave benchmark already exhibited
nontrivial angular structure. The mixed benchmark goes one step
further by combining this angular dependence with a nontrivial overall
scale set by both $s$- and $p$-wave interactions.  The hierarchy of
the three models shows how the benchmark improves as one supplements
the finite-volume dynamical signal with progressively more explicit
auxiliary information.

\section{Summary and Outlook}
\label{sec:summary}

The main motivation of this work is to demonstrate that
infinite-volume scattering information can be obtained from
finite-volume real-time evolution data and to show that machine
learning can be a useful component of scattering calculations on
quantum devices.

We therefore developed a first benchmark that connects finite-volume
time evolution to infinite-volume scattering information in a setting
that is simple enough to analyze thoroughly and rich enough to support
a realistic picture of potential issues.  We considered a
two-dimensional periodic lattice Hamiltonian for two distinguishable
particles with simple contact interactions and carried out real-time
wave-packet evolution and {\it measured} detector observables defined
by direct sums over lattice points. We calculated infinite-volume
total and differential cross sections and combined these with the
matching finite volume data.

We considered three separate scenarios: pure $s$-wave interactions, pure
$p$-wave interactions and the mixed case of $s+p$ wave
interactions. For these three scenarios we trained convolutional
neural networks and demonstrated that the neural networks are able to
{\it predict} infinite volume cross sections from finite volume data.

The present study is restricted to two distinguishable particles in
two spatial dimensions.  Furthermore, the Hamiltonians we have
considered here are intentionally simple: the angular infinite-volume
differential cross section is relatively simple, and the nontrivial
task is to infer the momentum-dependent infinite-volume signal from
finite-volume time-dependent data. Even so, our results establish this
approach as a first benchmark application of machine learning to
real-time scattering from finite-volume evolution, and provide
motivation for extending it to more complex systems.

This is only the first stage of a broader program: We will enlarge the
mixed-Hamiltonian scans further and furthermore study how the number
of {\it detector measurements} can be minimized while keeping the
neural networks predictive. We will furthermore go beyond the present
$s{+}p$ benchmark to more general operator families and work towards
implementing realistic nucleon-nucleon scattering in three spatial
dimensions.

One very important issue to resolve is also how to treat compound systems. In
particular, the case of several final state channels will complicate
the analysis significantly. A combination of several approaches, e.g. combining the results of
Ref.~\cite{Rule:2026brk} along with machine learning, may ultimately
be the most measurement effective implementation for the simulation of
scattering/reaction processes on quantum devices.

The results of this work should therefore be viewed as the foundation
for a more realistic scattering program aimed at finite-volume quantum
simulation and quantum computing applications. A key advantage of the
present approach for quantum computing applications is that the
detector observables require only standard-basis measurements, without
ancilla registers. Moreover, because the mapping from finite-volume
wedge-time signals to infinite-volume cross sections depends on the
universal low-energy structure of short-range scattering rather than
on the specific form of the Hamiltonian, a network trained on a
sufficiently diverse set of short-range interactions could in
principle serve as a general-purpose decoder applicable to new
Hamiltonians without retraining. Demonstrating this generalization is
a central goal of the next stage of this program.

\begin{acknowledgements}
  We are grateful to Adrian Del Maestro for useful discussions and to
  Thomas Papenbrock for comments on the manuscript.  AI-based tools
  were used to assist the software development and computational
  workflow. All scientific results were validated by the author. This
  work was supported by the National Science Foundation under Grant
  No.~PHY-2412612 and the US Department of Energy under Contract
  Nos.~DE-AC05-00OR22725 and DE-SC0021642.
\end{acknowledgements}


\begin{thebibliography}{19}%
\makeatletter
\providecommand \@ifxundefined [1]{%
 \@ifx{#1\undefined}
}%
\providecommand \@ifnum [1]{%
 \ifnum #1\expandafter \@firstoftwo
 \else \expandafter \@secondoftwo
 \fi
}%
\providecommand \@ifx [1]{%
 \ifx #1\expandafter \@firstoftwo
 \else \expandafter \@secondoftwo
 \fi
}%
\providecommand \natexlab [1]{#1}%
\providecommand \enquote  [1]{``#1''}%
\providecommand \bibnamefont  [1]{#1}%
\providecommand \bibfnamefont [1]{#1}%
\providecommand \citenamefont [1]{#1}%
\providecommand \href@noop [0]{\@secondoftwo}%
\providecommand \href [0]{\begingroup \@sanitize@url \@href}%
\providecommand \@href[1]{\@@startlink{#1}\@@href}%
\providecommand \@@href[1]{\endgroup#1\@@endlink}%
\providecommand \@sanitize@url [0]{\catcode `\\12\catcode `\$12\catcode
  `\&12\catcode `\#12\catcode `\^12\catcode `\_12\catcode `\%12\relax}%
\providecommand \@@startlink[1]{}%
\providecommand \@@endlink[0]{}%
\providecommand \url  [0]{\begingroup\@sanitize@url \@url }%
\providecommand \@url [1]{\endgroup\@href {#1}{\urlprefix }}%
\providecommand \urlprefix  [0]{URL }%
\providecommand \Eprint [0]{\href }%
\providecommand \doibase [0]{http://dx.doi.org/}%
\providecommand \selectlanguage [0]{\@gobble}%
\providecommand \bibinfo  [0]{\@secondoftwo}%
\providecommand \bibfield  [0]{\@secondoftwo}%
\providecommand \translation [1]{[#1]}%
\providecommand \BibitemOpen [0]{}%
\providecommand \bibitemStop [0]{}%
\providecommand \bibitemNoStop [0]{.\EOS\space}%
\providecommand \EOS [0]{\spacefactor3000\relax}%
\providecommand \BibitemShut  [1]{\csname bibitem#1\endcsname}%
\let\auto@bib@innerbib\@empty
\bibitem [{\citenamefont {Feynman}(1982)}]{Feynman:1981tf}%
  \BibitemOpen
  \bibfield  {author} {\bibinfo {author} {\bibfnamefont {R.~P.}\ \bibnamefont
  {Feynman}},\ }\href {\doibase 10.1007/BF02650179} {\bibfield  {journal}
  {\bibinfo  {journal} {Int. J. Theor. Phys.}\ }\textbf {\bibinfo {volume}
  {21}},\ \bibinfo {pages} {467} (\bibinfo {year} {1982})}\BibitemShut
  {NoStop}%
\bibitem [{\citenamefont {Jordan}\ \emph {et~al.}(2012)\citenamefont {Jordan},
  \citenamefont {Lee},\ and\ \citenamefont {Preskill}}]{Jordan:2012wd}%
  \BibitemOpen
  \bibfield  {author} {\bibinfo {author} {\bibfnamefont {S.~P.}\ \bibnamefont
  {Jordan}}, \bibinfo {author} {\bibfnamefont {K.~S.~M.}\ \bibnamefont {Lee}},
  \ and\ \bibinfo {author} {\bibfnamefont {J.}~\bibnamefont {Preskill}},\
  }\href {\doibase 10.1126/science.1217069} {\bibfield  {journal} {\bibinfo
  {journal} {Science}\ }\textbf {\bibinfo {volume} {336}},\ \bibinfo {pages}
  {1130} (\bibinfo {year} {2012})}\BibitemShut {NoStop}%
\bibitem [{\citenamefont {Simenel}(2012)}]{Simenel:2012ca}%
  \BibitemOpen
  \bibfield  {author} {\bibinfo {author} {\bibfnamefont {C.}~\bibnamefont
  {Simenel}},\ }\href {\doibase 10.1140/epja/i2012-12152-0} {\bibfield
  {journal} {\bibinfo  {journal} {Eur. Phys. J. A}\ }\textbf {\bibinfo {volume}
  {48}},\ \bibinfo {pages} {152} (\bibinfo {year} {2012})}\BibitemShut
  {NoStop}%
\bibitem [{\citenamefont {Kamada}\ \emph {et~al.}(2001)\citenamefont {Kamada}
  \emph {et~al.}}]{Kamada:2001tv}%
  \BibitemOpen
  \bibfield  {author} {\bibinfo {author} {\bibfnamefont {H.}~\bibnamefont
  {Kamada}} \emph {et~al.},\ }\href {\doibase 10.1103/PhysRevC.64.044001}
  {\bibfield  {journal} {\bibinfo  {journal} {Phys. Rev. C}\ }\textbf {\bibinfo
  {volume} {64}},\ \bibinfo {pages} {044001} (\bibinfo {year} {2001})},\
  \Eprint {http://arxiv.org/abs/nucl-th/0104057} {arXiv:nucl-th/0104057}
  \BibitemShut {NoStop}%
\bibitem [{\citenamefont {Deltuva}\ and\ \citenamefont
  {Fonseca}(2009)}]{Deltuva:2009zz}%
  \BibitemOpen
  \bibfield  {author} {\bibinfo {author} {\bibfnamefont {A.}~\bibnamefont
  {Deltuva}}\ and\ \bibinfo {author} {\bibfnamefont {A.~C.}\ \bibnamefont
  {Fonseca}},\ }\href {\doibase 10.1103/PhysRevC.79.014606} {\bibfield
  {journal} {\bibinfo  {journal} {Phys. Rev. C}\ }\textbf {\bibinfo {volume}
  {79}},\ \bibinfo {pages} {014606} (\bibinfo {year} {2009})}\BibitemShut
  {NoStop}%
\bibitem [{\citenamefont {Gloeckle}\ \emph {et~al.}(1996)\citenamefont
  {Gloeckle}, \citenamefont {Witala}, \citenamefont {Huber}, \citenamefont
  {Kamada},\ and\ \citenamefont {Golak}}]{Gloeckle:1996zz}%
  \BibitemOpen
  \bibfield  {author} {\bibinfo {author} {\bibfnamefont {W.}~\bibnamefont
  {Gloeckle}}, \bibinfo {author} {\bibfnamefont {H.}~\bibnamefont {Witala}},
  \bibinfo {author} {\bibfnamefont {D.}~\bibnamefont {Huber}}, \bibinfo
  {author} {\bibfnamefont {H.}~\bibnamefont {Kamada}}, \ and\ \bibinfo {author}
  {\bibfnamefont {J.}~\bibnamefont {Golak}},\ }\href {\doibase
  10.1016/0370-1573(95)00085-2} {\bibfield  {journal} {\bibinfo  {journal}
  {Phys. Rept.}\ }\textbf {\bibinfo {volume} {274}},\ \bibinfo {pages} {107}
  (\bibinfo {year} {1996})}\BibitemShut {NoStop}%
\bibitem [{\citenamefont {L{\"u}scher}(1986{\natexlab{a}})}]{Luscher:1985dn}%
  \BibitemOpen
  \bibfield  {author} {\bibinfo {author} {\bibfnamefont {M.}~\bibnamefont
  {L{\"u}scher}},\ }\href@noop {} {\bibfield  {journal} {\bibinfo  {journal}
  {Commun. Math. Phys.}\ }\textbf {\bibinfo {volume} {104}},\ \bibinfo {pages}
  {177} (\bibinfo {year} {1986}{\natexlab{a}})}\BibitemShut {NoStop}%
\bibitem [{\citenamefont {L{\"u}scher}(1986{\natexlab{b}})}]{Luscher:1986pf}%
  \BibitemOpen
  \bibfield  {author} {\bibinfo {author} {\bibfnamefont {M.}~\bibnamefont
  {L{\"u}scher}},\ }\href {\doibase 10.1007/BF01211097} {\bibfield  {journal}
  {\bibinfo  {journal} {Commun. Math. Phys.}\ }\textbf {\bibinfo {volume}
  {105}},\ \bibinfo {pages} {153} (\bibinfo {year}
  {1986}{\natexlab{b}})}\BibitemShut {NoStop}%
\bibitem [{\citenamefont {Burbano}\ \emph {et~al.}(2025)\citenamefont
  {Burbano}, \citenamefont {Carrillo}, \citenamefont {Urek}, \citenamefont
  {Ciavarella},\ and\ \citenamefont {Brice{\~n}o}}]{Burbano:2025reso}%
  \BibitemOpen
  \bibfield  {author} {\bibinfo {author} {\bibfnamefont {D.~A.}\ \bibnamefont
  {Burbano}}, \bibinfo {author} {\bibfnamefont {M.~A.}\ \bibnamefont
  {Carrillo}}, \bibinfo {author} {\bibfnamefont {R.}~\bibnamefont {Urek}},
  \bibinfo {author} {\bibfnamefont {A.~N.}\ \bibnamefont {Ciavarella}}, \ and\
  \bibinfo {author} {\bibfnamefont {R.~A.}\ \bibnamefont {Brice{\~n}o}},\
  }\href@noop {} {\bibfield  {journal} {\bibinfo  {journal} {arXiv e-prints}\
  ,\ \bibinfo {pages} {arXiv:2506.06511}} (\bibinfo {year} {2025})},\ \Eprint
  {http://arxiv.org/abs/2506.06511} {arXiv:2506.06511 [hep-lat]} \BibitemShut
  {NoStop}%
\bibitem [{\citenamefont {Sharma}\ \emph {et~al.}(2024)\citenamefont {Sharma},
  \citenamefont {Papenbrock},\ and\ \citenamefont {Platter}}]{Sharma:2024bjs}%
  \BibitemOpen
  \bibfield  {author} {\bibinfo {author} {\bibfnamefont {S.}~\bibnamefont
  {Sharma}}, \bibinfo {author} {\bibfnamefont {T.}~\bibnamefont {Papenbrock}},
  \ and\ \bibinfo {author} {\bibfnamefont {L.}~\bibnamefont {Platter}},\ }\href
  {\doibase 10.1103/PhysRevC.109.L061001} {\bibfield  {journal} {\bibinfo
  {journal} {Phys. Rev. C}\ }\textbf {\bibinfo {volume} {109}},\ \bibinfo
  {pages} {L061001} (\bibinfo {year} {2024})}\BibitemShut {NoStop}%
\bibitem [{\citenamefont {Turro}\ \emph {et~al.}(2024)\citenamefont {Turro},
  \citenamefont {Wendt}, \citenamefont {Quaglioni}, \citenamefont {Pederiva},\
  and\ \citenamefont {Roggero}}]{Turro:2024vpc}%
  \BibitemOpen
  \bibfield  {author} {\bibinfo {author} {\bibfnamefont {F.}~\bibnamefont
  {Turro}}, \bibinfo {author} {\bibfnamefont {K.~A.}\ \bibnamefont {Wendt}},
  \bibinfo {author} {\bibfnamefont {S.}~\bibnamefont {Quaglioni}}, \bibinfo
  {author} {\bibfnamefont {F.}~\bibnamefont {Pederiva}}, \ and\ \bibinfo
  {author} {\bibfnamefont {A.}~\bibnamefont {Roggero}},\ }\href {\doibase
  10.1103/PhysRevC.110.054604} {\bibfield  {journal} {\bibinfo  {journal}
  {Phys. Rev. C}\ }\textbf {\bibinfo {volume} {110}},\ \bibinfo {pages}
  {054604} (\bibinfo {year} {2024})}\BibitemShut {NoStop}%
\bibitem [{\citenamefont {Yusf}\ \emph {et~al.}(2025)\citenamefont {Yusf},
  \citenamefont {Gan}, \citenamefont {Moffat},\ and\ \citenamefont
  {Rupak}}]{Yusf:2025eqc}%
  \BibitemOpen
  \bibfield  {author} {\bibinfo {author} {\bibfnamefont {M.}~\bibnamefont
  {Yusf}}, \bibinfo {author} {\bibfnamefont {L.}~\bibnamefont {Gan}}, \bibinfo
  {author} {\bibfnamefont {C.}~\bibnamefont {Moffat}}, \ and\ \bibinfo {author}
  {\bibfnamefont {G.}~\bibnamefont {Rupak}},\ }\href {\doibase
  10.1103/PhysRevC.111.034001} {\bibfield  {journal} {\bibinfo  {journal}
  {Phys. Rev. C}\ }\textbf {\bibinfo {volume} {111}},\ \bibinfo {pages}
  {034001} (\bibinfo {year} {2025})}\BibitemShut {NoStop}%
\bibitem [{\citenamefont {Bennewitz}\ \emph {et~al.}(2025)\citenamefont
  {Bennewitz}, \citenamefont {Ware}, \citenamefont {Schuckert}, \citenamefont
  {Lerose}, \citenamefont {Muzzio}, \citenamefont {Halimeh}, \citenamefont
  {Surace}, \citenamefont {Pichler},\ and\ \citenamefont
  {Knap}}]{Bennewitz:2025mss}%
  \BibitemOpen
  \bibfield  {author} {\bibinfo {author} {\bibfnamefont {E.}~\bibnamefont
  {Bennewitz}}, \bibinfo {author} {\bibfnamefont {B.}~\bibnamefont {Ware}},
  \bibinfo {author} {\bibfnamefont {A.}~\bibnamefont {Schuckert}}, \bibinfo
  {author} {\bibfnamefont {A.}~\bibnamefont {Lerose}}, \bibinfo {author}
  {\bibfnamefont {F.}~\bibnamefont {Muzzio}}, \bibinfo {author} {\bibfnamefont
  {J.~C.}\ \bibnamefont {Halimeh}}, \bibinfo {author} {\bibfnamefont {F.~M.}\
  \bibnamefont {Surace}}, \bibinfo {author} {\bibfnamefont {H.}~\bibnamefont
  {Pichler}}, \ and\ \bibinfo {author} {\bibfnamefont {M.}~\bibnamefont
  {Knap}},\ }\href {\doibase 10.22331/q-2025-06-17-1773} {\bibfield  {journal}
  {\bibinfo  {journal} {Quantum}\ }\textbf {\bibinfo {volume} {9}},\ \bibinfo
  {pages} {1773} (\bibinfo {year} {2025})}\BibitemShut {NoStop}%
\bibitem [{\citenamefont {Taylor}(1972)}]{Taylor:1972pty}%
  \BibitemOpen
  \bibfield  {author} {\bibinfo {author} {\bibfnamefont {J.~R.}\ \bibnamefont
  {Taylor}},\ }\href@noop {} {\emph {\bibinfo {title} {{Scattering Theory: The
  Quantum Theory of Nonrelativistic Collisions}}}}\ (\bibinfo  {publisher}
  {John Wiley {\&} Sons, Inc.},\ \bibinfo {address} {New York},\ \bibinfo
  {year} {1972})\BibitemShut {NoStop}%
\bibitem [{\citenamefont {Chai}\ \emph {et~al.}(2025)\citenamefont {Chai},
  \citenamefont {Crippa}, \citenamefont {Jansen}, \citenamefont {K{\"u}hn},
  \citenamefont {Pascuzzi}, \citenamefont {Tacchino},\ and\ \citenamefont
  {Tavernelli}}]{Chai:2025fwp}%
  \BibitemOpen
  \bibfield  {author} {\bibinfo {author} {\bibfnamefont {Y.}~\bibnamefont
  {Chai}}, \bibinfo {author} {\bibfnamefont {A.}~\bibnamefont {Crippa}},
  \bibinfo {author} {\bibfnamefont {K.}~\bibnamefont {Jansen}}, \bibinfo
  {author} {\bibfnamefont {S.}~\bibnamefont {K{\"u}hn}}, \bibinfo {author}
  {\bibfnamefont {V.~R.}\ \bibnamefont {Pascuzzi}}, \bibinfo {author}
  {\bibfnamefont {F.}~\bibnamefont {Tacchino}}, \ and\ \bibinfo {author}
  {\bibfnamefont {I.}~\bibnamefont {Tavernelli}},\ }\href {\doibase
  10.22331/q-2025-02-19-1638} {\bibfield  {journal} {\bibinfo  {journal}
  {Quantum}\ }\textbf {\bibinfo {volume} {9}},\ \bibinfo {pages} {1638}
  (\bibinfo {year} {2025})}\BibitemShut {NoStop}%
\bibitem [{\citenamefont {Davoudi}\ \emph {et~al.}(2024)\citenamefont
  {Davoudi}, \citenamefont {Hsieh},\ and\ \citenamefont
  {Kadam}}]{Davoudi:2024swh}%
  \BibitemOpen
  \bibfield  {author} {\bibinfo {author} {\bibfnamefont {Z.}~\bibnamefont
  {Davoudi}}, \bibinfo {author} {\bibfnamefont {C.-C.}\ \bibnamefont {Hsieh}},
  \ and\ \bibinfo {author} {\bibfnamefont {S.~V.}\ \bibnamefont {Kadam}},\
  }\href {\doibase 10.22331/q-2024-11-11-1520} {\bibfield  {journal} {\bibinfo
  {journal} {Quantum}\ }\textbf {\bibinfo {volume} {8}},\ \bibinfo {pages}
  {1520} (\bibinfo {year} {2024})}\BibitemShut {NoStop}%
\bibitem [{\citenamefont {Rule}\ and\ \citenamefont
  {Stetcu}(2026)}]{Rule:2026brk}%
  \BibitemOpen
  \bibfield  {author} {\bibinfo {author} {\bibfnamefont {E.}~\bibnamefont
  {Rule}}\ and\ \bibinfo {author} {\bibfnamefont {I.}~\bibnamefont {Stetcu}},\
  }\href@noop {} {\  (\bibinfo {year} {2026})},\ \Eprint
  {http://arxiv.org/abs/2603.26881} {arXiv:2603.26881 [nucl-th]} \BibitemShut
  {NoStop}%
\bibitem [{\citenamefont {Adhikari}(1986)}]{Adhikari:1986}%
  \BibitemOpen
  \bibfield  {author} {\bibinfo {author} {\bibfnamefont {S.~K.}\ \bibnamefont
  {Adhikari}},\ }\href {\doibase 10.1119/1.14623} {\bibfield  {journal}
  {\bibinfo  {journal} {Am. J. Phys.}\ }\textbf {\bibinfo {volume} {54}},\
  \bibinfo {pages} {362} (\bibinfo {year} {1986})}\BibitemShut {NoStop}%
\bibitem [{\citenamefont {Chadan}\ \emph {et~al.}(1998)\citenamefont {Chadan},
  \citenamefont {Khuri}, \citenamefont {Martin},\ and\ \citenamefont
  {Wu}}]{Chadan:1998kq}%
  \BibitemOpen
  \bibfield  {author} {\bibinfo {author} {\bibfnamefont {K.}~\bibnamefont
  {Chadan}}, \bibinfo {author} {\bibfnamefont {N.~N.}\ \bibnamefont {Khuri}},
  \bibinfo {author} {\bibfnamefont {A.}~\bibnamefont {Martin}}, \ and\ \bibinfo
  {author} {\bibfnamefont {T.~T.}\ \bibnamefont {Wu}},\ }\href {\doibase
  10.1103/PhysRevD.58.025014} {\bibfield  {journal} {\bibinfo  {journal} {Phys.
  Rev. D}\ }\textbf {\bibinfo {volume} {58}},\ \bibinfo {pages} {025014}
  (\bibinfo {year} {1998})}\BibitemShut {NoStop}%
\end{thebibliography}
\end{document}